# EVALUATING PRODUCTION PLANNING AND CONTROL SYSTEMS IN DIFFERENT ENVIRONMENTS: A COMPARATIVE SIMULATION STUDY

Wolfgang Seiringer[1], Balwin Bokor[1], and Klaus Altendorfer[1]

[1]Dept. of Production Operation Management, University of Applied Sciences Upper Austria,
Steyr, Upper Austria, OOE, AUSTRIA

## ABSTRACT

Selecting the appropriate production planning and control systems (PPCS) presents a significant challenge for many companies, as their performance, i.e., overall costs, depends on the production system environment. Key environmental characteristics include the system's structure, i.e., flow shop, hybrid shop, or job shop, and the planned shop load. Besides selecting a suitable PPCS, its parameterization significantly influences the performance. This publication investigates the performance and the optimal parametrization of Material Requirement Planning (MRP), Reorder Point System (RPS), and Constant Work In Progress (ConWIP) at different stochastic multi-item multi-stage production system environments by conducting a comprehensive full factorial simulation study. The results indicate that MRP and ConWIP generally outperform RPS in all observed environments. Moreover, when comparing MRP with ConWIP, the performance clearly varies depending on the specific production system environment.

## 1 INTRODUCTION

The main goal of production planning is to ensure that the production systems' output precisely aligns with customer demand, serving as a critical bridge between operational capabilities and market expectations. This objective places production planning and control at the core of manufacturing companies, embodying both challenging and important roles. The challenge arises from the necessity to manage a tremendous amount of complex information, such as customer demand, Bill of Materials (BoM), work plans, and more (Reuter et al. 2017). Meanwhile, their importance is highlighted by their impact on the production system performance, i.e., overall costs (Hopp and Spearman 2011).

To address the complexities of information and enhance efficiency, various production planning and control systems (PPCS) have been developed. Among these, Material Requirements Planning (MRP), Reorder Point System (RPS), and Constant Work In Progress (ConWIP) are widely recognized in research and frequently adopted in practice. Selecting the most suitable PPCS presents a significant challenge for many companies due to the unique nature of their production system environments. A critical factor in this decision-making process is the production system environment, encompassing aspects such as the production system structure or the planned shop load. Particularly, the structure of the production system, whether it is a flow shop, hybrid shop, or job shop can significantly impact the performance of a PPCS. In a flow shop, production orders move through a series of sequential workstations. Conversely, in a job shop the processing sequence is tailored to the specific requirements of the production order. A hybrid shop merges aspects of both, enabling flexible operation sequences at certain workstations while others follow a predetermined order (Hillier et al. 1999).

Transitioning from the discussion on the critical role of production system environments on PPCS performance – despite the extensive research conducted on these systems – the comparative analysis remains relatively scarce. Gupta and Snyder (2009) identified only 20 articles comparing two or more PPCS. Since the research of Gupta and Snyder (2009), only three significant studies comparing PPCS emerged: Jodlbauer and Huber (2008), Miclo et al. (2019), and Thürer et al. (2022). Jodlbauer and Huber (2008) evaluated MRP, Kanban, ConWIP and Drum-Buffer-Rope (DBR), focusing on parameter stability and environmental robustness for a multi-item multi-stage flow shop production system. They found ConWIP to be superior when optimally parametrized, but also noted its lack of robustness, as minor deviations from optimal parametrization led to significant performance deviations. Moreover, they highlighted the robustness of MRP against environmental uncertainty and mentioned the necessity for additional parametrization of Kanban in case environmental uncertainty diminishes. Miclo et al. (2019) compared Demand Driven MRP (DDMRP) with MRP and Kanban within a flexible multi-item multi-stage flow shop production system, taking into account different levels of demand uncertainty. Their simulation study led to the conclusion that DDMRP outperformed both methods, while MRP was found to be the least effective, regardless of the level of demand uncertainty. Thürer et al. (2022) adapted the model of Jodlbauer and Huber (2008) by integrating a single bottleneck and different due date tightness, in addition to neglecting stochastic aspects, i.e., machine breakdowns, scrap parts, and lot sizes, to compare MRP, Kanban, DBR, and DDMRP. Their findings emphasized the superiority of DBR and DDMRP, especially over MRP. Moreover, they highlighted that tighter due dates require PPCS that realize shorter production lead times, i.e., time span from actual production start to actual production end. Yet, comprehensive comparisons of different PPCS remain scarce, and the findings are largely inconclusive. Additionally, the examination of various environmental characteristics, especially the performance impact of different production system structures, is still lacking. This gap is particularly critical as real-world production systems continue to increase in complexity (Bergmann and Heinicke 2017).

Therefore, this publication conducts a comprehensive full factorial simulation study to evaluate the performance of MRP, RPS, and ConWIP across different stochastic multi-item and multi-stage production system environments. In doing so, the authors explore three distinct production system structures: flow shop, hybrid shop, and job shop, alongside three levels of planned shop load. The performance is evaluated based on various cost components, including WIP costs, finished goods inventory (FGI) costs and tardiness costs. Moreover, the study also discusses the optimal parameterization of each PPCS at different production system environments. Thus, the following research questions are addressed:

- RQ1: Which production planning and control system (MRP, RPS, ConWIP) demonstrates superior performance across diverse production system structures (flow shop, hybrid shop, job shop) and different planned shop loads?
- RQ2: How do environmental characteristics, especially production system structures and planned shop load, necessitate adjustments in the parametrization of MRP, RPS, and ConWIP for optimal performance?

This research offers valuable insights for both the academic field and managerial practice. Academically, it contributes to addressing the scarce research on PPCS comparisons and explores the research gap in evaluating PPCS across various environmental characteristics, focusing on production system structures. Managerially, it provides decision-makers with a detailed analysis of the most effective PPCS method under specific environments and investigates PPCS performance as well as the approximated optimal parameterization for different environments.

This publication is structured as follows: Section 2 provides a brief overview of PPCS characteristics and delves into the operational specifics of MRP, RPS, and ConWIP. In Section 3, the complex simulation model is introduced and the three production system structures observed. The comprehensive numerical study is outlined in Section 4, followed by a discussion of the results in Section 5. The publication concludes with final thoughts and suggestions for further research.

## 2 PRODUCTION PLANNING AND CONTROL SYSTEMS

To provide an overview of the investigated PPCS, first, four key characteristics that allow for differentiation are detailed. Subsequently, these characteristics, as well as the production planning and order release mechanisms, together with the planning parameters for the three investigated PPCS, are explored. Lastly, the PPCS are summarized based on the outlined characteristics.

## 2.1 Characteristics

Firstly, PPCS can be classified based on their *operational mechanism* as either *push* or *pull*. The literature offers numerous definitions for these mechanisms. The authors align upon the definition provided by Hopp and Spearman (2004), where *pull* systems restrict the amount of WIP within the production system, in contrast to *push* systems, which do not impose an explicit limit on WIP. Secondly, PPCS are differentiated by their *demand orientation*, distinguishing between *demand-driven* systems and those *authorizing production*. *Demand-driven* systems leverage information on future demands to plan production, whereas systems that *authorize production* rely on downstream demand or customer withdrawal (Cochran and Kaylani 2008). Thirdly, the *control structure* is identified as either *centralized* or *decentralized*. *Centralized* systems rely on a single authorization unit for production planning and order release, whereas *decentralized* systems distribute decision-making to individual units on the shop floor level (Woschank et al. 2021). Lastly, the planning complexity is evaluated, reflected in the *number of planning parameters*, distinguishing between *system-level parameterization* and *item-level parameterization*. *System-level parameterization* applies universally across all items, whereas *item-level parameterization* involves setting parameters specifically for each item, respective component. This distinction is crucial, as *system-level parameterization* significantly reduces the effort needed for maintaining accurate master data, thereby ensuring performance (Pansara 2023).

## 2.2 Material Requirements Planning

Material Requirements Planning (MRP) is a *push* PPCS developed by Orlicky (1975), where production planning and order release are based on four *centrally* controlled steps. These four steps are: netting, lot-sizing, backward scheduling, and BoM explosion also described in detail by Hopp and Spearman (2011). At the netting step, material quantities are determined by offsetting gross requirements – derived from customer orders and forecasts – against the current inventory, excluding safety stock to prevent depletion during planning, and incorporating scheduled receipts (Matsuura and Tsubone 1991). Given that MRP leverages information concerning customer orders and forecasts, MRP can be characterized as a *demand-driven* PPCS. At the lot-sizing step, the net requirements can be batched based on lot-sizing policies to balance set-up and ordering effort against inventory holding (Yelle 1979). Two commonly applied lot-sizing policies for MRP are Fixed Order Quantity (FOQ) and Fixed Order Period (FOP). FOQ orders a predetermined quantity or a multiple of it upon each time reordering occurs, whereas FOP batches net requirements within predefined time intervals. At the next step, planned start dates are established by backward scheduling from the planned end date based on the planned lead times. Lastly at the BoM explosion, the steps are repeated for the underlying MRP item, systematically to the deepest BoM level. As the WIP is not explicitly restricted by performing these steps, MRP is a *push* PPCS. To perform these four steps, three *item-level* planning parameters are required: safety stock, lot size with the chosen lot-sizing policy, and planned lead time. The safety stock is essential for buffering against shortages due to unexpected demands with short customer required lead times or scrap, yet it raises inventory costs. The lot-size aims to balance set-up and ordering effort against inventory holding, influencing setup frequency, machine occupation per batch and shop load. The planned lead time defines the available time to produce the respective item, including waiting times due to machine occupation and considering the inherited fluctuation of the production system, i.e., stochastic processing time (Altendorfer 2019).

## 2.3 Reorder Point System

The Reorder Point System (RPS) is a *pull* PPCS, leveraging on the Economic Order Quantity model, which focuses on minimizing the inventory management costs, i.e., set-up and ordering effort and inventory holding, by employing optimized order quantities at stock replenishment (Silver et al. 1998). Thereby, production planning and order release are *decentralized* for each item based on comparing the reorder point with the inventory position (Hopp and Spearman 2011). The inventory position of an item is determined by its current inventory level plus any scheduled receipts minus any backorders. Thus, RPS *authorizes* *production* without leveraging demand information. If the inventory position falls below the reorder point, a production order is scheduled and released based on FOQ lot-sizing policy, i.e., a predetermined quantity or a multiple of it (Seiringer et al. 2023). Therefore, the planning parameters are the reorder point and the lot size on an *item-level* basis, where the reorder point must meet the demand during the replenishment time and consider uncertainties in demand and production. Additionally, RPS has a conceptual connection to the well-known PPCS Kanban, as discussed by Yang (1998). Kanban is considered a subtype of RPS, where the reorder point in Kanban is effectively the sum of all Kanban containers minus one, multiplied by the container lot size. However, it distinguishes itself with its card system to initiate new production orders. This inherent relationship implies that the study, while focusing on RPS, also covers Kanban principles and applications.

### 2.4 Constant Work in Progress

Spearman et al. (1990) introduced Constant Work in Progress (ConWIP) as a *pull* alternative to Kanban, where production planning is based on a work-ahead window and order release is controlled through cards by associating WIP and WIP-cap with production orders. Setting the WIP-cap based on production orders lead to a constrained ability for load balancing. Addressing this limitation, Thürer et al. (2019) linked the WIP-cap to workload measured in standard processing time, i.e., required time to process the production order, rather than on the count of production orders. This linkage improved the performance significantly. Building on this, for further discussion, WIP and WIP-cap are also associated with workload, aligning with the approach to enhance performance. In detailing production planning, ConWIP is *demand-driven* as net requirements are determined based on either a Master Production Schedule (MPS) or directly from on customer orders. The MPS can apply lot-sizing policies at the creation of the production orders to balance set-up and ordering effort against inventory holding. Production order release is only permitted if the due date falls within the work-ahead window extended by the current date, effectively acting as a scheduling window. Thus, the work-ahead window prevents premature production order release, thereby controlling FGI (Bokor and Altendorfer 2024). These production orders are prioritized according to Earliest Due Date (EDD), with *centralized* release permitted only in case the shop floor WIP is below the WIP-cap. The released production orders are then dispatched at conventional ConWIP based on First-In-System-First-Out (FISFO) (Spearman et al. 1990). Both planning parameters, i.e., work-ahead window and WIP-cap, apply universally across all items, offering a significant advantage through *system-level parameterization* (Spearman et al. 2022). However, as noted by Jaegler et al. (2018), complex production system structures might necessitate additional, independently parameterized ConWIP-loops to sustain performance.

Table 1 summarizes the four characteristics for the three investigated PPCS. To underscore the significance of planning complexity, particularly relevant in simulation studies where combinatorics result in a vast array of combinations during full factorial enumeration, a summary of the number of planning parameters is also included. As stated by Law (2014), full factorial designs are highly valued for their thoroughness and the depth of insight they provide. However, they can become resource-intensive and time-consuming with an increase in the number of factors and levels, as the total number of experiments (simulations) increases exponentially (Seiringer et al. 2022). Here, *n* represents the number of items or components, highlighting how an increase in the number of items significantly expands the number of possible combinations and thereby the scope and scale of the analysis.

Table 1: Characteristics of investigated PPCS.

| | MRP | RPS | ConWIP |
|---|---|---|---|
| Operational mechanism | push | pull | pull |
| Demand orientation | demand-driven | authorizing production | demand-driven |
| Control structure | centralized | decentralized | centralized |
| Planning complexity | item-level | item-level | system-level |
| Required planning parameters | 3n | 2n | 2 |

## 3 SIMULATION MODEL

To assess the performance of the three investigated PPCS, a stochastic multi-item multi-stage simulation model was developed, integrating three different production system structures, focusing on flow shop, hybrid shop, and job shop. First, the investigated production system structures, including the BoM, are outlined. Afterwards, the customer demand, including the customer required lead time and the connection to the planned shop load, is discussed. This is followed by an in-depth exploration of the integrated production planning and order release mechanisms of the observed PPCS, building upon the foundational concepts presented in Section 2. Lastly, the order processing in the simulation model is discussed.

### 3.1 Production System Structure

Each production system structure consists of eight items, each with a specific share of the total demand, represented by the proportions {0.100, 0.075, 0.200, 0.125, 0.075, 0.150, 0.150, 0.125} for the $i^{th}$ item, i.e., for the first item 0.100; and so on. Each item, as well as components at lower levels, requires just one component from the preceding level. However, to process an item or component, also more machines can be required at one BoM level. The lowest BoM level at each production system structure corresponds to the item, whereas the highest level of the BoM level represents the raw material, which is always available and not planned. Figure 1 offers a detailed overview of the three observed production system structures, including their BoM configurations.

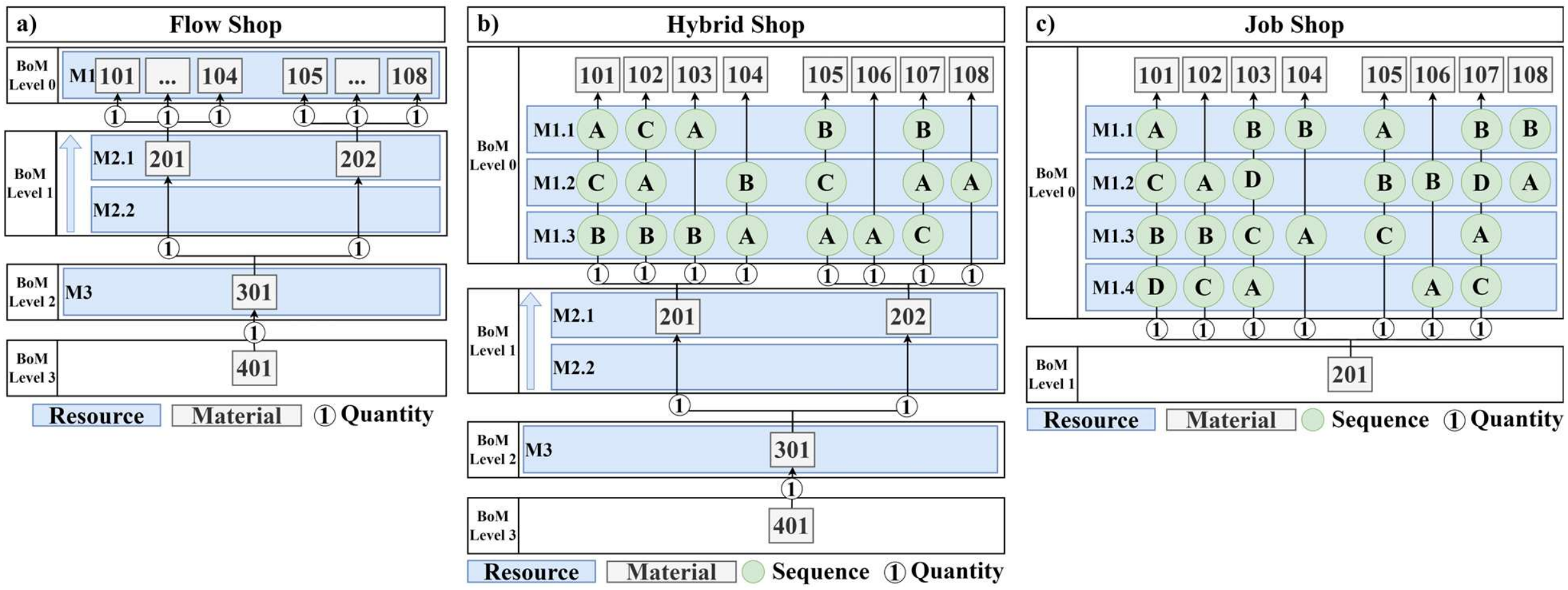

Figure 1: Investigated Production System Structures including Bill of Material: a) flow shop; b) hybrid shop; c) job shop

As depicted in Figure 1 a), the flow shop structure is designed with four machines and extends through four BoM levels, diverging at both BoM Level 2 and BoM Level 1 as it approaches the final item. Moreover, at BoM Level 1, two machines are required to produce the components, i.e., component 201 and 202. Figure 1 b) introduces the hybrid shop structure, which incorporates six machines and maintains the same four BoM levels with a similar divergence. It initiates with a flow shop material flow but transitions into a job shop production system at BoM Level 0, where the production path is determined by the item and not all machines are necessary for processing. The sequence of machines at BoM Level 0 is indicated by green-bordered letters, starting with 'A' and so forth. Lastly, the job shop structure, as depicted in Figure 1 c), involves four machines and two BoM levels. Since the BoM Level 1 at the job shop structure corresponds to the raw material, only one BoM level (BoM Level 0) is planned. This BoM Level 0 is similar to the identical BoM level at the hybrid shop structure, as the sequence of the machines varies and not all machines are required to process each item.

### 3.2 Customer Demand

To assess the influence of planned shop load levels on PPCS performance, the expected mean customer order quantities of the eight items are adjusted, resulting in planned shop loads of 85 %, 90 %, and 95 %, represented as ∈ {0.85, 0.90, 0.95}. Nonetheless, the proportionality of the items remains constant as stated in Section 3.1. Despite varying planned shop loads, a deterministic customer order for each item, containing just that single item, is generated in every period. To include uncertainty in the customer demand, the actual quantities of these eight items follow the lognormal distribution with a coefficient of variation (CV) of 0.2. The customer required lead time consists of a fixed portion of ten periods and a lognormal distributed variable proportion with an expected mean of five periods and a CV of 0.5.

### 3.3 Production Planning and Order Release

The production planning and order release mechanism differs for each PPCS and is generally discussed in Section 2. To specify for this publication, concerning MRP, the FOP lot-sizing policy is applied, and each component, e.g., BoM Level 0, is planned separately. Nevertheless, to mitigate the complexity of combinatorics as detailed in Table 1, the planning parameters are standardized across all components and separately across all items. Thus, for instance, the components 201, 202, and 301 depicted in Figure 1 share identical planning parameters. For RPS, the standard configuration is applied, and again distinctions concerning the planning parameters are only made between components and items. For ConWIP, an MPS is integrated that batches gross requirements based on FOP lot-sizing policy to balance set-up effort against inventory holding. Moreover, in scenarios involving more than one planned BoM level, such as in the flow shop and hybrid shop production structures, two ConWIP loops are implemented to enhance system performance, as described by Huang et al. (2015). One loop is designated for items, e.g., BoM Level 0, while the other one manages all components. With two ConWIP loops in place, the planned start dates for items are determined by backward scheduling based on the estimated item lead time, and the planned start date for components is set by further backward scheduling from this point, based on the estimated component lead time. The earliest planned start date for items is calculated by subtracting the work-ahead-window buffer from the planned start dates for items, facilitating earlier release in case components are available. Therefore, the work-ahead window equals the estimated item lead time plus the work-ahead-window buffer. Upon completion of a production order at the last machine in a ConWIP loop, the WIP level for that specific ConWIP loop is reduced by the workload, i.e., standard processing time including set-up, associated with the production order.

### 3.4 Order Processing

The released orders are processed based on their required production path. The expected mean processing time varies for each item or component at each machine. The expected mean setup time within one machine is not varied, accounting for 10 % of the available production time. This results in reduced setup times as the number of items or components processed on a single machine increase, and vice versa. The expected required capacity, i.e., the time to produce demand including setup, as well as the available capacity, are identical across all machines. This uniformity ensures a consistently planned shop load for each machine, regardless of the production system structure. Hence, no bottleneck machine exists. Both the actual processing time and actual set-up time follow a lognormal distribution with a CV of 0.2. Since MRP and RPS do not mandate a specific dispatching rule, First-In-First-Out (FIFO) is utilized, while for ConWIP, FISFO is implemented, following its conventional application as specified by Spearman et al. (1990). After completion, the finished goods remain in the FGI until the customer required due date is reached. In case of tardiness, the delivery is executed immediately. The released orders are processed based on their required production path. The expected mean processing time varies for each item respective component at each machine.

In summary, the stochastic aspects of the simulation involve both customer demand and order processing. For customer demand, there are two different stochastic influences: the calculation of the actual quantities for each order and the customer-required lead time. In order processing, two stochastic influences are considered: the calculation of the actual processing time and the actual setup time. All stochastic variables are derived from lognormal distributions. The log-normal distribution was selected due to its non-negativity and its widespread application in previous studies.

# 4 NUMERICAL STUDY

To comprehensively explore the performance of the investigated PPCS, a full factorial simulation study is conducted for nine different production system environments. Table 2 summarizes all tested production system environments as well as planning parameters used in the full factorial enumeration. The planning parameter ranges were determined based on preliminary studies.

Table 2: Investigated environments and planning parameters for each PPCS.

| | | Min | Max | Step size | Iterations |
|---|---|---|---|---|---|
| Env. | Production system structure | - | - | - | 3 |
| | Planned shop load | 0.85 | 0.95 | 0.05 | 3 |
| Different production system environments | | | | | 9 |
| MRP | Planned lead time items [days] | 1 | 6 | 1 | 6 |
| | FOP lot size items [days] | 1 | 4 | 1 | 4 |
| | Safety stock items [prop. demand] | 0 | 1.5 | 0.5 | 4 |
| | Planned lead time components [days] | 1 | 3 | 1 | 3 |
| | FOP lot size components [days] | 1 | 4 | 1 | 4 |
| | Safety stock components [prop. item demand] | 0 | 1.5 | 0.5 | 4 |
| | Total iterations MRP for {flow shop; hybrid; job shop} production system environment | | | | {13,824; 13,824; 288} |
| RPS | Reorder-point items [prop. demand] | 3 | 7 | 0.5 | 9 |
| | FOQ lot size items [prop. demand] | 0.5 | 3 | 0.5 | 6 |
| | Reorder-point components [prop. item demand] | 1 | 4 | 0.5 | 7 |
| | FOQ lot size components [prop. item demand] | 0.5 | 3 | 0.5 | 6 |
| | Total iterations ROP for {flow shop; hybrid; job shop} production system environment | | | | {6,804; 6,804; 162} |
| ConWIP | MPS FOQ lot size [prop. demand] | 1 | 3 | 0.5 | 5 |
| | WIP-cap item / component [workload in minutes] | 10,000 | 50,000 | 10,000 | 5 |
| | Estimated lead time items [days] | 1 | 5 | 1 | 5 |
| | Estimated lead time components [days] | 1 | 5 | 1 | 5 |
| | Work-ahead-window buffer [days] | 0 | 3 | 1 | 4 |
| | Total iterations ConWIP for{flow shop; hybrid; job shop} production system environment | | | | {7,500; 7,500; 375} |
| **Total iterations / simulation runs for all PPCS and production system environments** | | | | | **57,081 / 570,810** |

Initially, the production system environments are categorized by their production system structure, i.e., flow shop, hybrid shop, and job shop. Each structure is analyzed at three levels of planned shop load, creating nine unique environments. For each environment, the optimal planning parameters for all three PPCS are approximated. For MRP, the planned lead time in days is set, the FOP lot-sizing policy is applied, measured in days, and safety stock levels are set as a proportion of the item's expected demand per day, or for components, as the sum of the demands of items requiring that component. For example, setting a safety stock level of two for an item with an expected demand of 50 per day results in a total safety stock of 100. For RPS: the reorder point and the lot size are set based on a similar ratio-based logic. For ConWIP, an MPS is applied, which batches gross requirements based on the FOQ lot-sizing policy, separate ConWIP loops are applied for items and components, whereby WIP and WIP-cap are associated with workload in minutes and both WIP-caps are set identically, an estimated item lead time, measured in days, is applied for backward scheduling in the case of two ConWIP loops, a work-ahead-window buffer, also measured in days, is used to determine the earliest start date for items by subtracting the work-ahead-window buffer from the planned start date, and in cases with only one ConWIP loop, the estimated item lead time serves as the work-ahead window.As highlighted in Section 3.3, to reduce the combinatorial effect associated with setting the planning parameters in a full factorial enumeration for MRP and RPS, identical planning parameters are employed across all items or components within a single production system environment. Since the job shop production system structure features only one planned BoM level, only the planning parameters concerning the items are planned, which also results in a single ConWIP-loop.

For MRP, the authors conduct 13,824 iterations for the flow and hybrid shop, and 288 for the job shop. For RPS, there are 6,804 iterations for the flow and hybrid shop, and 162 for the job shop. For ConWIP, 7,500 iterations are performed for the flow and hybrid shop, with 375 for the job shop. To ensure the robustness of the results derived from the stochastic simulation model, ten replications per iteration are performed. This results in a total of 570,810 simulation runs. Each replication lasts 400 days with a 150-day warm-up phase. Parallel computing is used across 21 computers with Intel Core i5-10500 CPUs, each with six cores and 32 GB memory. Each computer sequentially selects simulation parameters from a simulation server utilizing a PostgreSQL relational database, executes the simulation, and then delivers the results back to the server. The simulation model was developed using AnyLogic, the parameter combinations are generated using RStudio, and the results are stored in the same PostgreSQL database.

## 5 NUMERICAL RESULTS

The performance of the three observed PPCS, i.e., MRP, RPS and ConWIP, across the nine production system environments is evaluated based on overall costs, including WIP costs, finished goods inventory (FGI) costs and tardiness costs. By doing so, the minimum overall costs of all tested parameter combinations are identified. All cost components are calculated as the mean value of the ten replications conducted. Storing FGI is more costly compared to storing components due to their higher value and the additional handling and storage requirements needed to maintain their quality. This is also true for (WIP) items, leading to FGI and WIP inventory costs being twice that of components. Specifically, the cost structure is as follows: 0.5 cost units (CU) per day for WIP components, 1 CU per day for inventory components, 1 CU per day for WIP items, 2 CU per day for inventory items, and 38 CU per item per day for tardiness. The tardiness costs of 38 CU per item and day represent the costs of stockouts, based on a target service level of 95 %, calculated using the formula: $service\ level = 1 - (FGI\ costs)/(FGI\ costs + tardiness\ costs)$ (refer to the papers in Axsäter (2015) for more details). For simplicity, transportation costs for intralogistics material handling are not considered. The WIP components refer to the work-in-progress of unfinished goods, which includes all BoM levels except Level 0. In contrast, WIP items refer to the work-in-progress for finished goods, hence regarding BoM Level 0. At first, the performance of the three observed PPCS across the nine different production system environments is evaluated. Subsequently, the optimal parameterization, i.e., the planning parameters, of each PPCS necessary to achieve this performance is discussed.

### 5.1 Performance

Figure 2 visualizes the overall costs per day, including a detailed presentation of the cost components for the three observed PPCS at the nine different production system environments, resulting from the three planned shop loads within three production system structures. As the Figure implies, RPS is less effective than both MRP and ConWIP across all nine observed production environments. This inferior performance is attributed to the higher FGI costs needed to maintain low tardiness costs, arising from a lack of demand information utilization and strictly authorized production. This observation is consistent with Schonberger and Schniederjans (1984), who highlighted the necessity for high inventory levels in traditional inventory control methods compared to approaches that leverage demand information. Comparing ConWIP and MRP, ConWIP demonstrates superior performance in flow and hybrid shop production systems with planned shop loads of 0.85 and 0.90. However, at a planned shop load of 0.95, MRP outperforms ConWIP in these production system structures. These findings closely correspond with Jodlbauer and Huber (2008), who noted the superiority of ConWIP over MRP and Kanban in flow shop production systems, the latter of which is conceptually connected to RPS as discussed in Section 2.3. In the job shop production system, MRP generally surpasses ConWIP, except at the highest planned shop load of 0.95, where ConWIP achieves significantly lower tardiness costs.

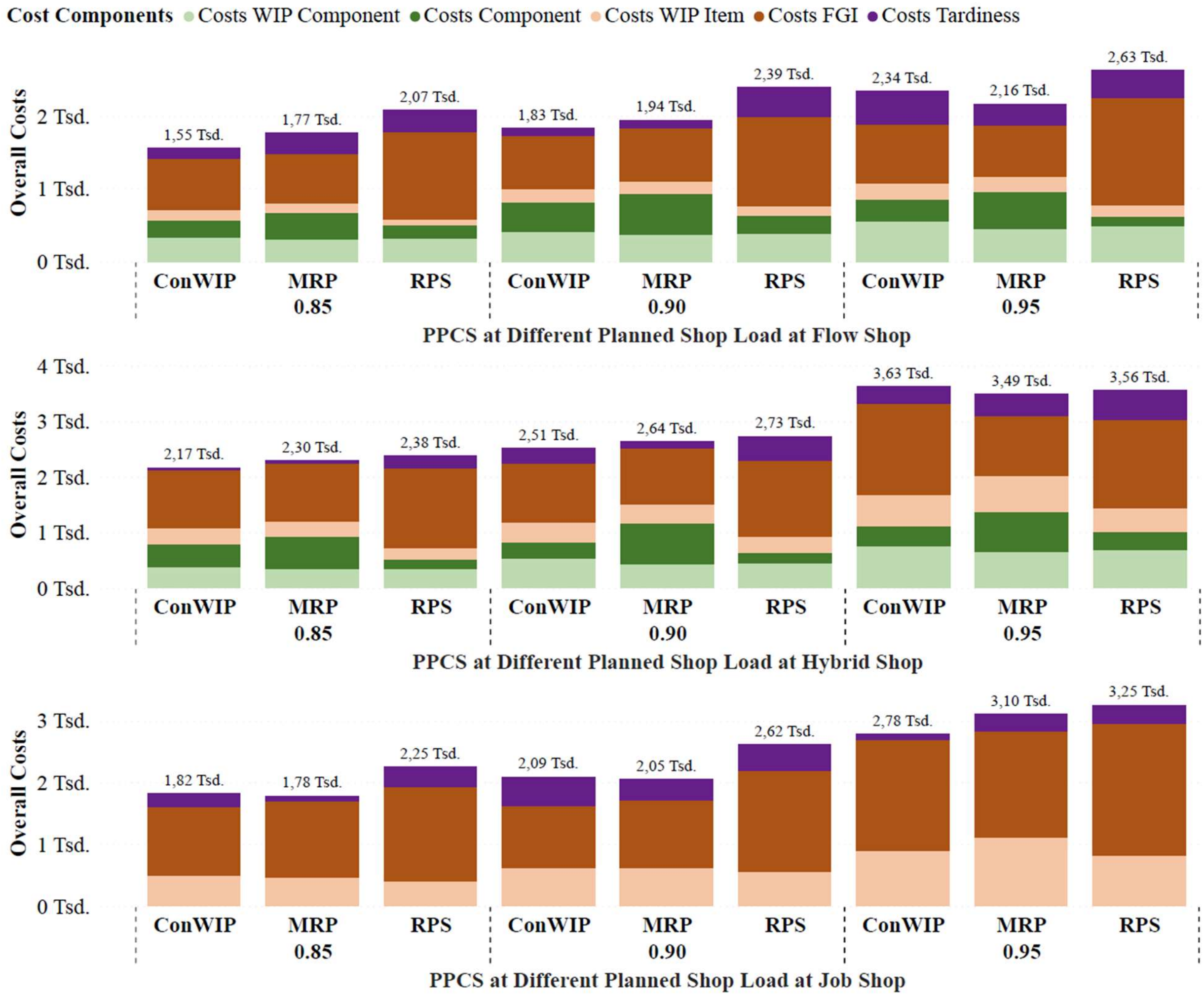

Figure 2: Overall costs per day of investigated PPCS.

## 5.2 Optimal Planning Parameters

Table 3 shows the optimal planning parameters, i.e., lead to the minimum overall costs, for the three PPCS across the nine observed production system environments. Observing MRP reveals that in eight out of nine observed production system environments safety stocks for both components and items are required. Moreover, a higher planned shop load necessitates either a greater safety stock or a longer planned lead time, evident in the hybrid and job shop production systems. The increase in safety stocks or longer planned lead times absorbs the higher workload to protect against potential tardiness. Concerning the planned lead time, this is particularly evident at the 0.95 planned shop load in the job shop production system. These findings are consistent with the research presented by Altendorfer (2019). Moreover, MRP implementation in the job shop production system necessitates the longest planned lead time. This requirement is partially due to the modeled job shop production system structure, where the BoM level includes the highest number of machines (up to four), as depicted in Figure 1 c). Additionally, within this job shop production system structure, a higher planned shop load leads to increased lot sizes, a trend that is uniquely observed in this system.

Table 3: Optimal planning parameters at different environments for each PPCS.

| | | | | | | | | | | |
|---|---|---|---|---|---|---|---|---|---|---|
| Env. | Production system structure | Flow Shop | | | Hybrid Shop | | | Job Shop | | |
| | Planned shop load | 0.85 | 0.90 | 0.95 | 0.85 | 0.90 | 0.95 | 0.85 | 0.90 | 0.95 |
| MRP | Planned lead time items [days] | 1 | 1 | 1 | 2 | 2 | 3 | 3 | 3 | 5 |
| | FOP lot size items [days] | 1 | 1 | 1 | 1 | 1 | 1 | 1 | 1 | 2 |
| | Safety stock items [prop. demand] | 0.5 | 0.5 | 0.5 | 0.5 | 0.5 | 0.5 | 0.5 | 0.5 | 0.5 |
| | Planned lead time components [days] | 2 | 2 | 2 | 2 | 2 | 3 | - | - | - |
| | FOP lot size components [days] | 1 | 1 | 1 | 1 | 1 | 1 | - | - | - |
| | Safety stock components [prop. item demand] | 0.0 | 0.5 | 0.5 | 0.5 | 1 | 0.5 | - | - | - |
| RPS | Reorder-point items [prop. demand] | 3.0 | 3.0 | 3.5 | 4.0 | 4.0 | 4.5 | 4.5 | 5.0 | 6.0 |
| | FOQ lot size items [prop. demand] | 1.0 | 1.0 | 1.0 | 1.0 | 1.0 | 1.0 | 1.0 | 1.0 | 2.0 |
| | Reorder-point components [prop. item demand] | 1.0 | 1.0 | 1.0 | 1.0 | 1.0 | 1.5 | - | - | - |
| | FOQ lot size components [prop. item demand] | 1.0 | 1.0 | 1.0 | 1.0 | 1.0 | 1.0 | - | - | - |
| ConWIP | MPS FOQ lot size [prop. demand] | 1.0 | 1.0 | 1.0 | 1.0 | 1.0 | 3.0 | 1.0 | 1.0 | 1.0 |
| | WIP-cap item / component [workload in Tsd. minutes] | 20 | ≥ 20 | ≥ 30 | ≥ 20 | ≥ 30 | 30 | 20 | ≥ 30 | 20 |
| | Estimated lead time items [days] | 1 | 1 | 1 | 2 | 2 | 3 | 3 | 3 | 5 |
| | Estimated lead time components [days] | 3 | 4 | 4 | 4 | 4 | 5 | - | - | - |
| | Work-ahead-window [days] | ≥ 1 | ≥ 1 | ≥ 1 | ≥ 1 | ≥ 1 | ≥ 1 | - | - | - |

Observing RPS, higher planned shop loads lead to increased reorder points for items, e.g., FGI, significantly contributing to the inferior performance discussed in Section 5.1. This increase is due to longer replenishment times as machines are often busy with production orders, resulting in more waiting times. However, the reorder points for components remain largely unaffected in both flow shop and hybrid shop systems, even with higher planned loads. Yang (1998) also found a link between planned shop loads and increased reorder points in a single machine system. The lack of impact on reorder points at the BoM levels, such as components, is a new insight. Additionally, lot sizes increase only in the job shop system at the highest planned load, similar to MRP behavior.

For ConWIP, as discussed in Section 3.3, in a single ConWIP-loop scenario like the job shop system, estimated lead times for items act as the work-ahead window. In other production systems, start and end dates are determined through backward scheduling using estimated lead times for items or components. In a two ConWIP loops scenario, the work-ahead window equals the estimated lead time plus the work-ahead-window buffer. As with MRP, higher planned loads result in longer lead times, extending the work-ahead window, especially in the job shop system, as noted by Bokor and Altendorfer (2024). The estimated lead times for items match MRP planned lead times, but for components, they are much longer than for MRP. This is due to the fixed MPS lot sizes in ConWIP and the absence of a safety buffer, necessitating longer lead times to account for uncertainty. In ConWIP, the job shop system does not require larger MPS lot sizes at higher planned shop loads, while a 0.95 planned shop load in the hybrid system significantly increases the optimal lot size. Lastly, increasing the WIP-cap and work-ahead-window buffer beyond a certain point neither improves nor worsens performance significantly. This is particularly noticeable with the work-ahead-window buffer: increasing it from zero to one enhances performance, but higher values do not yield further benefits, since the components are not stocked any earlier.

## 6 CONCLUSION

This article evaluated the performance of three PPCS (MRP, RPS, and ConWIP) across nine stochastic multi-item multi-stage production environments through a full factorial simulation. Performance was evaluated on WIP costs, FGI costs, and tardiness costs. The authors developed a simulation model, integrating the PPCS into flow shop, hybrid shop, and job shop structures, each under three planned shop loads. They computed optimal planning parameters for each PPCS across all environments by minimizing these costs. Concerning the performance comparison, two key findings emerge: 1) Results indicate a superior performance of MRP and ConWIP over RPS across all nine observed scenarios, stemming from RPS's lack of demand information utilization and strictly authorized production, leading to higher FGI to maintain low tardiness costs. 2) When comparing ConWIP and MRP, ConWIP exhibits superior performance at 0.85 and 0.90 planned shop loads in the flow and hybrid shop production systems. However, MRP outperforms ConWIP at these planned shop loads in the job shop production system.

Regarding the impact of various production system environments on the optimal planning parameters for each PPCS, three observations are identified for each of them. For MRP, 1) safety stock is nearly always required for both components and items, 2) higher planned shop loads necessitate increased safety stocks or planned lead times, and 3) the job shop production systems require the longest planned lead times as well as an increased lot size at higher planned shop load. For RPS, 1) higher planned shop loads result in increased reorder points for FGI as replenishment times increase due to occupied machines, negatively impacting performance, 2) reorder points for components remain stable regardless of shop load in both the flow and hybrid shop production systems, and 3) similar to MRP, lot sizes only increase in the job shop production system at the highest planned shop load. For ConWIP, 1) in scenarios with two ConWIP loops, the estimated lead time of items is identical and exhibits similar behavior as the planned lead time of items in MRP. 2) conversely, in each observed environment, the estimated lead time of components is generally longer than the planned lead time of components in MRP, due to a consistent lot size based on the MPS and the absence of a safety stock, and 3) unlike in MRP, the job shop production system under ConWIP does not require increased lot sizes at higher planned shop loads, though the hybrid production system structure does. Further research should explore additional production system environments, potentially including variables such as different customer required lead times or the impact of machine breakdowns. Comparisons with less-known PPCS, such as DBR or Demand Driven MRP, should also be considered.

## ACKNOWLEDGMENTS

This research was funded in whole, or in part, by the Austrian Science Fund (FWF) [P 32954-G]. For the purpose of open access, the author has applied a CC BY public copyright license to any Author Accepted Manuscript version arising from this submission.

## AUTHOR BIOGRAPHIES

**WOLFGANG SEIRINGER** is a Research Associate in the field of Operations Managemen at the University of Applied Sciences Upper Austria and Ph.D. candidate at the JKU Linz for Logistics and Supply Management. Email: wolfgang.seiringer@fh-steyr.at.

**BALWIN BOKOR** is a Research Associate in the field of Operations Management at the University of Applied Sciences Upper Austria and Ph.D. candidate at the JKU Linz for Logistics and Supply Management. Email: balwin.bokor@fh-steyr.at.

**KLAUS ALTENDORFER** is a Professor of Operations Management at the University of Applied Sciences Upper Austria, expert in production system simulation, stochastic inventory model, and production planning. Email: klaus.altendorfer@fh-steyr.at.